\documentclass[sigconf]{acmart}

\AtBeginDocument{%
  \providecommand\BibTeX{{%
    \normalfont B\kern-0.5em{\scshape i\kern-0.25em b}\kern-0.8em\TeX}}}

\setcopyright{none}
\copyrightyear{2024}
\acmYear{2024}
\acmDOI{XXXXXXX.XXXXXXX}

\acmConference[Preprint - SIGCSE'24]{Technical Symposium on Computer Science Education (Accepted)}{March 20--25,
  2024}{Portland, OR}
\acmPrice{XX.00}
\acmISBN{978-1-4503-XXXX-X/18/06}

\usepackage{listings}
\usepackage{xcolor}
\usepackage{pgfplots}
\usepackage{pgfplotstable}

\pgfplotsset{width=\linewidth,compat=1.9}

\definecolor{backcolour}{rgb}{0.95,0.95,0.92}
\lstdefinestyle{mystyle}{
    backgroundcolor=\color{backcolour},   
    commentstyle=\color{codegreen},
    keywordstyle=\color{magenta},
    numberstyle=\tiny\color{codegray},
    stringstyle=\color{codepurple},
    basicstyle=\ttfamily\footnotesize,
    breakatwhitespace=false,         
    breaklines=true,                 
    captionpos=b,                    
    keepspaces=true,                 
    numbers=none,             
    numbersep=5pt,                  
    showspaces=false,                
    showstringspaces=false,
    showtabs=false,                  
    tabsize=2
}

\graphicspath{{graphics}}

\begin{document}

\title[dcc \texttt{-{}-}help: Generating Context-Aware Compiler Error Explanations with Large Language Models]{dcc \texttt{-{}-}help: Transforming the Role of the Compiler by Generating Context-Aware Error Explanations with Large Language Models}

\author{Andrew Taylor}
\affiliation{%
  \institution{University of New South Wales}
  \streetaddress{2052}
  \city{Sydney}
  \state{New South Wales, 2052}
  \country{Australia}}

\author{Alexandra Vassar}
\affiliation{%
  \institution{University of New South Wales}
  \streetaddress{2052}
  \city{Sydney}
  \state{New South Wales, 2052}
  \country{Australia}}
  
\author{Jake Renzella}
\affiliation{%
  \institution{University of New South Wales}
  \streetaddress{2052}
  \city{Sydney}
  \state{New South Wales, 2052}
  \country{Australia}}

\author{Hammond Pearce}
\affiliation{%
  \institution{University of New South Wales}
  \streetaddress{2052}
  \city{Sydney}
  \state{New South Wales, 2052}
  \country{Australia}}

\newcommand{\numusers}{2,565~}
\newcommand{\numuses}{64,451~}

\begin{abstract}
In the challenging field of introductory programming, high enrolments and failure rates drive us to explore tools and systems to enhance student outcomes, especially automated tools that scale to large cohorts. This paper presents and evaluates the \emph{dcc \texttt{-{}-}help} tool, an integration of a Large Language Model (LLM) into the Debugging C Compiler (DCC) to generate unique, novice-focused explanations tailored to each error. \emph{dcc \texttt{-{}-}help} prompts an LLM with contextual information of compile- and run-time error occurrences, including the source code, error location and standard compiler error message. The LLM is instructed to generate novice-focused, actionable error explanations and guidance, designed to help students understand and resolve problems without providing solutions. \emph{dcc \texttt{-{}-}help} was deployed to our CS1 and CS2 courses, with 2,565 students using the tool over 64,000 times in ten weeks. We analysed a subset of these error/explanation pairs to evaluate their properties, including conceptual correctness, relevancy, and overall quality. We found that the LLM-generated explanations were conceptually accurate in 90\% of compile-time and 75\% of run-time cases, but often disregarded the instruction not to provide solutions in code. Our findings, observations and reflections following deployment indicate that \emph{dcc \texttt{-{}-}help} provides novel opportunities for scaffolding students' introduction to programming.
\end{abstract}

\ccsdesc[300]{Software and its engineering~Compilers}
\ccsdesc[500]{Social and professional topics~CS1}
\ccsdesc[300]{Computing methodologies~Natural language processing}

\keywords{CS1, AI in CS1, AI in Education, Generative AI, Large Language Models, Compiler Error Messages, Debugging, Error Message Enhancement, Programming Error Messages}

\maketitle

\section{Introduction}\label{Introduction}

Programming has remained a difficult concept to teach and learn, with globally high attrition and failure rates in introductory computing (CS1) courses \cite{Bennedsen2019FailureLater}. One of the most common difficulties cited when learning to program is the inability to read, understand and act on compiler error messages \cite{Becker2016EffectiveStudents, Karvelas2020TheBehavior, Kohn2019ThePython}. While confusing compiler error messages may be frustrating for even an experienced programmer, in some cases, they may create a learning barrier for novices. Previously, the authors of the Debugging C Compiler (DCC) showed how DCC improved the viability of teaching C in introductory programming courses \cite{Taylor2023FoundationsCompiler}. The existing implementation of DCC, which is open-source and publicly available, produces enhanced compiler error messages and explanations at both compile- and run-time to support novices in addressing common C errors. The authors claim that the tool was in part motivated by growing enrolments in CS1 courses in the previous decade, with our institution's computing cohorts growing by 45\%, introducing challenges when providing adequate support to meet the demand at this scale.

While DCC's enhanced error detection and explanations assist students in writing safer, more correct C code, students can still require assistance from teaching teams to explain error messages in terms they understand. Increased cohort sizes mean productivity and motivation are impacted, as students frequently encounter delays in receiving these necessary explanations.

Large Language Models (LLMs) are a form of neural network; an artificial intelligence that can learn the context and meaning of written language, including code, from large datasets of text used to train models. The open release of one such model in November 2022, OpenAI's ChatGPT (\url{https://chat.openai.com}), has sparked interest in how educators can use these types of models to improve outcomes in CS1. 

While recent studies have evaluated the efficacy of using LLMs to generate compiler error explanations \cite{MacNeil2022GeneratingModel, Leinonen2023UsingMessages, Becker2023ProgrammingGeneration}, there has been no published integration of an LLM into the compiler itself. This could transform the role of the compiler from simply generating error messages to producing detailed, contextualised, natural language guidance and feedback designed for novices. In this work we contribute such a tool. It can support novice programmers at scale, providing bespoke, on-demand guidance to support their learning. Our tool is open source and can be found at \url{http://dcc.cse.unsw.edu.au}.

\section{Background}\label{Background}

\subsection{Compiler Error Messages}
Compilers are a primary interface between a novice programming student and learning a language; however, issues interpreting and acting on compiler error messages have been well documented \cite{Barik2014CompilerNotifications, Becker2016EffectiveStudents, Karvelas2020TheBehavior, Prather2017OnApproach, Traver2010OnMean}. Students have even been known to change their majors, citing cryptic and hard-to-learn compiler error messages as one of the reasons \cite{Denny2021OnFactors}. Initial work has been done to enhance the readability of compiler error messages and explore the use of enhanced error messages \cite{Becker2019UnexpectedFuture, Becker2016EffectiveStudents, Becker2018TheTest, Carvalho2021EnhancedCS1, Denny2014EnhancingIneffectual, Pettit2017DoInconclusive}. However, the results of this work are inconclusive, with no clear evidence in favour of enhanced compiler error messages \cite{Becker2019CompilerResearch}. Although there is evidence that suggests students are reading compiler error messages, it is not directly clear how many students successfully understand and act on them \cite{Becker2016EffectiveStudents, Denny2014EnhancingIneffectual}. One of the difficulties when enhancing error messages is the manual process involved in identifying and integrating the enhanced messages into the compiler. The hand-crafted nature of these explanations means that they fail to cover the breadth of possibilities of potential student errors. %

\subsection{The Debugging C Compiler}\label{dcc}
The Debugging C Compiler (DCC) is a C/C++ compiler designed for novice programming students \cite{Taylor2023FoundationsCompiler}. The tool has been used millions of times, by thousands of students, to enable a fundamentals-first introductory programming course \cite{Taylor2023FoundationsCompiler}. DCC supports students by providing enhanced compiler error messages, which detect common errors that standard C implementations (such as GCC and Clang) miss. DCC achieves its goals via the following features, all
incorporated into a single easy-to-use package  \cite{Taylor2023FoundationsCompiler}: 
\begin{itemize}
    \item \textbf{Additional compile- and run-time error detection}: DCC embeds run-time error detection tools, such as Valgrind \cite{NicholasNethercote2007Valgrind}, AddressSanitizer and GDB into the generated executable to provide additional information to the DCC error explanation system including call stack printout and memory leak detection. Clang and GCC static analysis options are used to provide additional compile-time checks.
    \item \textbf{Enhanced error messages}: The DCC explainer, both at compile-time and run-time, interprets and explains the most common novice error messages using simple, hand-crafted explanations for a range of common error types.
    \item \textbf{Additional context}: DCC embeds source code into the executable to illustrate the location of run-time errors. This provides additional information to the student and allows more efficient bug identification and resolution.
\end{itemize}

In \autoref{listing:runtime}, we compile and execute a program which attempts to access an uninitialised variable. GCC does not detect the error; however, DCC flags the bug, and the location of the error. 

\begin{lstlisting}[caption={Uninitialised Variable---\texttt{gcc}: no error, \texttt{dcc}: error.}, label = listing:runtime]
$ gcc program.c && ./a.out
0
$ dcc program.c && ./a.out
Runtime error: uninitialized variable accessed.
Execution stopped in main() in the program.c at line 6:

int main(void) { 
    int numbers[10];
    for (int i = 1; i < 10; i++) {
        numbers[i] = i;
    }
--> printf("%d\n", numbers[0]);
}
Values when execution stopped:

numbers = {<uninitialized value>,1,2,3,4,5,6,7,8,9}
numbers[0] = <uninitialized value>
\end{lstlisting}

DCC's access to additional context at run-time including original program source, error location (line number), and the GDB call stack at the moment a run-time error occurs is critically important to generate actionable, contextual explanations at run-time.

\subsection{Large Language Models and Education}\label{llms}
Large Language Models, such as GPT-3 and later Codex models~\cite{chen_evaluating_2021}, display significant general-purpose and cross-domain capabilities in natural language processing. These transformer-based models are trained over large quantities of text scraped from the internet, and in Codex's case, with code mined from millions of open-source repositories. Codex demonstrated state-of-the-art capabilities in code authorship via code-writing benchmark tests (such as HumanEval~\cite{chen_evaluating_2021}), later underpinned the commercial GitHub Copilot. %

A recent training methodology, Reinforcement Learning with Human Feedback (RLHF), can be applied to LLMs to produce models capable of following user intents~\cite{ouyang_training_2022}. This is beneficial to override the default tendency of LLMs to simply act as a `smart autocomplete', and makes it easier to have the models actually `follow instructions'. 
The premier LLM in this space is OpenAI's ChatGPT, which was fine-tuned from GPT-3 and Codex to provide conversational-style instruction following completions. In software engineering, ChatGPT can thus be used to help translate and debug code and provide code explanations using natural language. 

The adoption of ChatGPT-style LLMs within education is currently mixed, with some individuals, schools, and systems forbidding generative AIs (e.g. in Australian primary and secondary schools~\cite{linton_chatgpt_2023}) and others moving towards targeted and mass utilisation. The primary concern stems from the potential for wide-scale academic misconduct, but proponents of the technology argue that careful usage will unlock new pedagogical tools and strategies. Kasneci et al. provide a comprehensive survey in this area~\cite{kasneci_chatgpt_2023}, finding that, for example, ChatGPT is already being used for educational methods such as generating tests, quizzes, and flashcards. %

Research into the benefits and pitfalls of using LLMs to support novice learners in CS1 is still in its infancy \cite{MacNeil2022GeneratingModel, MacNeil2022AutomaticallyModels, Denny2023ConversingLanguage, Jalil2023ChatGPTPerils, Becker2023ProgrammingGeneration, Finnie-Ansley2022TheProgramming}. The majority of the work has focused on testing how well these tools can solve a public repository of CS1 programming problems (usually in languages other than C); and to what extent natural language modifications or prompts can lead to the generation of successful solutions. For instance, one study showed that Codex can perform better than most novice students on code writing questions in CS1 courses, scoring in the 75th percentile, and generating multiple solutions to problems \cite{Finnie-Ansley2022TheProgramming}. Another study found, using 31 questions from a popular software testing textbook, that ChatGPT was able to respond to 77.5\% of questions and provide a correct answer in 55.6\% of cases (further prompting of the tool led to a slightly higher rate of correct answers and explanations) \cite{Jalil2023ChatGPTPerils}. Denny et al. found GitHub Copilot to be effective in solving standard introductory programming problems, successfully solving about half of 166 problem sets the on first attempt, and a further 60\% of the remaining problems with some natural language changes to the problem specification \cite{Denny2023ConversingLanguage}. Concurrently with this work, Harvard's CS50 has released an `AI chatbot'~\cite{dreibelbis_harvards_2023} which aims to help students find bugs in their programs and perform Q\&A over unfamiliarities or error messages. This tool is external to the compiler and implemented in their online platform, utilising a bespoke LLM instead of ChatGPT to minimise accidental over-help. We instead explore the incorporation of ChatGPT into the compiler directly, such that it may generate on-demand, novice-friendly explanations of compile- and run-time errors. 

As the complexity of the problem grows, so does the reliance on human input and prompting \cite{Austin2021ProgramModels}. There are further concerns that tools such as Codex can lead to over-reliance on programming tasks \cite{chen_evaluating_2021}, where students agree with LLM output even when it is incorrect. %
Some research has shown that producing explanations can reduce the over-reliance on the LLM models and improve overall decision-making \cite{Vasconcelos2023ExplanationsDecision-Making, Bansal2021DoesPerformance}; however, issues regarding student integrity and over-reliance persist. To produce prompts capable of asking the right questions, the user must be able to understand the problem they are experiencing in the first instance, which is not often the case with a novice programming student.

Recent work by Leinonen et al. \cite{Leinonen2023UsingMessages} explored the evaluation of Codex in producing enhanced programming error messages. The study selected a subset of Python error messages that were reported by students as being the least readable, and then the researchers produced code examples that would trigger these types of errors. One of the limitations of this work is the lack of an authentic classroom setting and the absence of programs written by the students themselves. The study evaluated a series of prompts designed to explain compiler errors and generate actionable fixes, and subsequently, the quality of the code fixes and error explanations generated in response to the prompts. Leinonen et al. found that error message explanations and proposed fixes require improvement before being introduced in CS1 due to students' over-reliance and trust in the correctness of the message explanation. Still, Leinonen et al. found that plain-language explanations of errors can decrease how threatening compiler error messages appear, and could be instrumental in improving learning outcomes for students at scale.

In our case, we intend for the generative explanations to guide students to understand compiler output, including DCC output, which may be confusing or lacking in contextual details. By providing a simple AI-generated explanation in the development environment, we hope to support student progress and understanding without requiring delays until staff are available to assist.

\section{A new toolflow: Generative help}
We introduce a new toolflow whereby DCC uses the OpenAI ChatGPT 3.5 API (model: \emph{gpt-3.5-turbo-0301}) at compile- and run-time to consume source code, error messages, and locations to generate contextual, novice-friendly error and warning explanations designed to augment typical compiler output. Using the OpenAI API at run-time is only possible with a tool like DCC, as typical C implementations such as GCC and Clang do not have access to the source code in the executable.

\begin{figure*}[ht]
    \includegraphics[width=\linewidth]{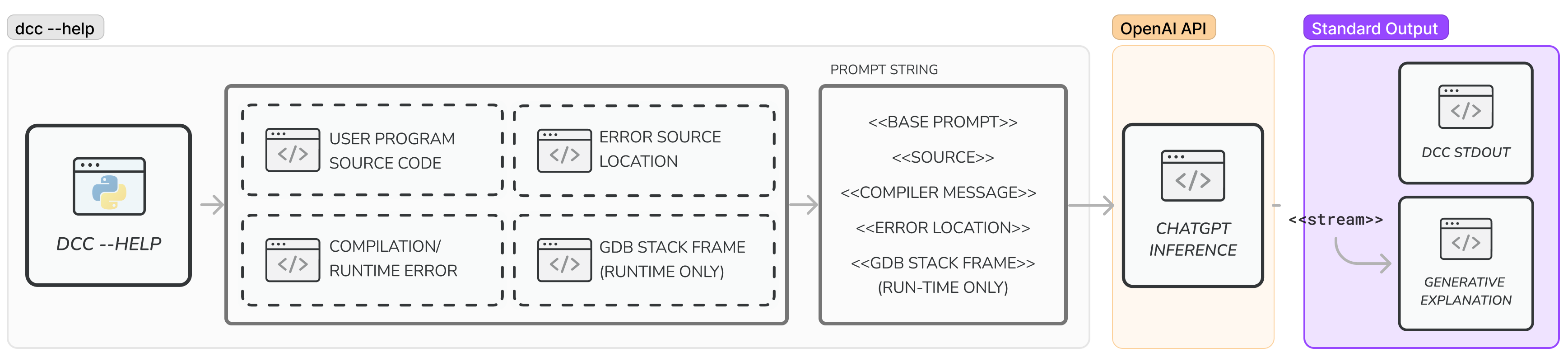}
    \caption{Diagram of New \textit{-{}-help} Generative Explanation Toolflow in the Debugging C Compiler at Compile- and Run-Time.}\label{fig1}
\end{figure*}

Our implementation was created by forking the open-source DCC Github repository and contributing the open-source \emph{-{}-}help extension (\url{https://github.com/COMP1511UNSW/dcc}). The overview of compile- and run-time process (\autoref{fig1}) describes how the new toolflow produces generative explanations. When \emph{dcc \texttt{-{}-}help} is executed (\autoref{listing:studenterr}), the previous error (compile- or run-time) is captured and prepared for an explanation by ChatGPT 3.5. A text prompt, not visible to the student, is generated (example of a prompt for run-time can be seen in ~\autoref{listing:apicall}), which includes:

\begin{itemize}
\item Base prompt,
\item Program source code,
\item DCC enhanced error message,
\item Error line number, and,
\item Values in the stack frame at error time (\textit{run-time only}).
\end{itemize}

\begin{lstlisting}[caption={Example of a Prompt Sent to ChatGPT, Not Visible by Users (Run-Time Error)}, label = listing:apicall]
system:content:
You are a tutor helping a student.
Do not fix the program.
Do not provide code.

user:content:
This is my C program <<Source Code>>
Help me understand this message from the C compiler:
<<DCC Enhanced Explanation>>
Error location: Line 6
Values: <<GDB Stack Frame for run-time>>
Remember, you are tutor helping a student.
Do not write code for the student.
\end{lstlisting}

The prompt, shown in \autoref{listing:apicall}, is sent via an HTTP API call to OpenAI's \emph{gpt-3.5-turbo-0301} and the response is streamed back into the student's terminal environment (\autoref{listing:studenterr}) in situ via the HTML Event stream format\footnote{\url{https://developer.mozilla.org/en-US/docs/Web/API/Server-sent_events}}. An example of the run-time error/explanation workflow explaining \autoref{listing:runtime} is shown in \autoref{listing:studenterr}.

\begin{lstlisting}[caption=Example Generative Explanation (Run-Time Error), label= listing:studenterr]
$ dcc program.c && ./a.out
<<Standard compiler output (compile-time only)>>
<<Non Gen-AI DCC enhanced explanation>>
Don't understand? Get AI-generated with `dcc --help`

$ dcc --help
Here is an AI generated explanation. Be careful - it may be wrong!

The error message you received is indicating that you are trying to access an uninitialized variable. In this case, the uninitialized variable is numbers[0].
In the program, you have declared an array called numbers which has 10 elements. However, you have not assigned any value to numbers[0] before trying to print its value with printf(). 
To fix this error, you can simply assign a value to numbers[0] before trying to print it. This can be done by adding numbers[0] = 0; before the printf() statement.
\end{lstlisting}

With \emph{dcc \texttt{-{}-}help}, we thus introduce a \emph{generative explanation} written by ChatGPT, in addition to the current DCC enhanced explanations. The complete nomenclature follows:
\begin{itemize}
    \item \textbf{Standard} compiler output for compile-time errors,
    \item \textbf{DCC enhanced error message} (\autoref{listing:studenterr}),
    \item \textbf{Generative explanation} (LLM-generated) (\autoref{listing:studenterr}). 
\end{itemize}

\subsection{Additional Design Choices}
Additional design decisions were made in order to safeguard the student learning experience. For example, \emph{dcc \texttt{-{}-}help} warns students that AI-generated explanations may not be correct. The tool also detects if a student is generating many \emph{dcc \texttt{-{}-}help} explanations in a short amount of time, warning them that they should use \emph{dcc \texttt{-{}-}help} sparingly, and should always understand the code they are writing---we do not make the AI help tool available in the exam environment. We present our reflections on these design choices in the discussion (\autoref{discussion}).

\section{Method}
This study aims to evaluate the efficacy of OpenAI's ChatGPT API (\emph{gpt-3.5-turbo-0301}) LLM in producing context-aware generative explanations in response to compiler errors in DCC. To evaluate the quality of responses, we released a version of DCC containing our help extension to students in a range of CS1 and CS2 courses at a large Australian university, then tracked its usage when generating error explanations. The CS1 curriculum at this institution covers a range of introductory topics, ranging from control flow to the use of arrays and linked lists. Students were informed that if they encountered either a compile- or run-time error, they could proceed to run the \emph{dcc \texttt{-{}-}help} command to execute the LLM inference and generate a response. We then logged each occurrence including the source code containing the error, the error location (line number), the raw C compiler error, and the ChatGPT response. The same prompt strategy was used in each instance, injecting the relevant source code, error location, and DCC error (\autoref{listing:apicall}). Additionally, usage statistics were collected by logging all student activities associated with the DCC and \emph{dcc \texttt{-{}-}help} tool.

\subsection{Data Extraction} 
This study utilises the student-generated error/explanation occurrences, including the source code of problematic code that arose throughout the students' regular coursework activities. We randomly sampled from the over 64,000 uses of \emph{dcc \texttt{-{}-}help} and extracted 200 compile-time errors and 200 run-time errors for a total evaluation set of 400 error/explanation pairs. An additional randomly selected mix of 15 error/explanation pairs was categorised jointly by the four reviewers together to reach consensus on categorisation strategies and to ensure a consistent approach. 

\subsection{Data Filtering}\label{filtering}
In line with the requirements of our relevant Ethics body's approval of this research project, data processing included anonymising the source code, including stripping potential student identifiers from comments and logs. Regular expressions were able to remove these from source code and filenames effectively, and best efforts were taken to remove names from comments (such as header-comments). Other comments, as they may describe the intended functionality of the code, were preserved---these source code comments will impact the quality of the LLM's responses. Finally, any staff who may have generated logs when testing the tool were filtered out.

\subsection{Data Analysis} \label{sssec:analysis}
Four reviewers (three authors of this paper and one student researcher) were each asked to evaluate 100 error/explanations from the extracted set of 400 as described in \autoref{filtering}. Each of the author reviewers has a significant history of teaching introductory computing, and the student researcher has been teaching introductory computing for the last two years. Reviewers were instructed to assess each LLM-generated explanation (student source code with error, compiler error message, and LLM-generated contextual explanation) across the following properties:

\begin{itemize}
    \item \textit{Conceptual accuracy} (Yes/No) - is the generated response conceptually correct? 
    \item \textit{Inaccuracy} (Yes/No) - are there inaccuracies present?
    \item \textit{Correctness} (Yes/No) - is the provided guidance technically correct resulting in being able to solve the problem?
    \item \textit{Relevance} (Yes/No) - is the generated message relevant to the encountered error?
    \item \textit{Completeness} (Yes/No) - is the provided explanation complete, not missing any critical information that would help students understand the error?
    \item \textit{Code Solution} (Yes/No) - is the solution provided as code in the generated response?
    \item \textit{Response type} (Peer/Tutor) - is the generated response commensurate in quality with a \textit{peer} or a \textit{tutor}? 
\end{itemize}

Prior to commencing the classification, the reviewers categorised 15 error/explanation pairs, separate from the analysis set, as a group to ensure that everyone had the same interpretation of the measured aspects. For each error/explanation pair, the reviewers could access the source code that generated the error, the output from DCC specifying the line number of where the error has occurred, the DCC enhanced explanation of the error, which also included the state of variables at the time when the error was produced (for run-time errors), and the LLM-generated explanation produced with this data. All of these were used in analysing the generated explanation by each reviewer. 

\subsection{Reliability of Evaluation}
Each reviewer was randomly assigned 100 errors/explanation pairs (50 compile-time, 50 run-time) for evaluation. In addition, ten percent of each reviewer's errors were randomly allocated to all other reviewers to determine inter-rater reliability, bringing the total errors analysed by each reviewer to 130. To address limitations of percentage agreement which does not take into account reviewers agreeing by chance, Light's Kappa \cite{Landis1977AnObservers} was used to determine an overall index of agreement.

\section{Results}
Results are presented in \autoref{table:1}. These provide the frequency of "Yes" responses across each category (see \autoref{sssec:analysis} for categories). In addition, the measure of inter-rater reliability across each category and the four reviewers is calculated using Light's kappa \cite{Landis1977AnObservers}. Guidelines \cite{Landis1977AnObservers} are also provided for interpreting the reliability values, where 0 indicates no agreement and 1 indicates perfect agreement; \emph{moderate} agreement is indicated by values 0.41 < $\kappa$ < 0.60, and 0.61 < $\kappa$ < 0.80 indicates \emph{substantial} agreement. Moderate agreement was observed for conceptual accuracy, the relevance of response, completeness of response, inaccuracy present, and type of response. Substantial agreement was observed in the technical correctness of the guidance. Overall, the LLM-generated explanations were clear across both compile-time and run-time errors, with better performance at compile-time in comparison to run-time. At compile-time, 90\% conceptual accuracy was observed, as compared to 75\% at run-time. No inaccuracy was present in the generative explanation for compile-time errors in 78\% of cases. Lower inaccuracy was observed at run-time, with only 53\% of explanations having no inaccuracy. This trend is observed in other categorisations also, specifically noting 93\% correctness at compile-time as compared to 66\% at run-time. 
Generative explanations were consistently deemed relevant to the error, with 92\% relevancy at compile-time and a reduced 75\% at run-time. There is a large difference in how complete the explanations are between compile-time explanations at 72\%, as compared to run-time, reduced 39\%. Despite the prompt instructing that no code be given out with the explanation, 48\% of the explanations at compile-time and 49\% of run-time explanations contained blocks of code that reviewers considered too much help. Finally, the generative explanations were approximated to an overall quality in terms of tutor-level or peer-level. Results found that explanations were deemed tutor-like for 72\% of compile-time explanations. In contrast, only 45\% of run-time explanations were deemed to be of a quality that tutors are expected to provide. Overall, run-time use of LLM-generated explanations was consistently worse. 

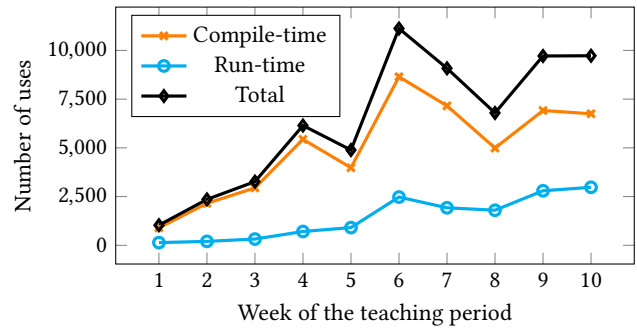
\begin{figure}[h]
\begin{tikzpicture}
\begin{axis}[%
xtick=data,
legend pos=north west,
ylabel = {Number of uses},
xlabel = {Week of the teaching period},
title={},
scaled y ticks = false,
        y tick label style={
            /pgf/number format/fixed,
            /pgf/number format/precision=1
        },
        height=5cm,
        ytick={0,2500,5000,7500,10000},
]

\addplot [mark=x, color=orange, very thick] table [x index = {0}, y index = {2}, col sep=comma] {data/stats.csv};
\addlegendentry{Compile-time};

\addplot [mark=o,color=cyan, very thick] table [x index = {0}, y index = {3}, col sep=comma] {data/stats.csv};
\addlegendentry{Run-time};

\addplot [mark=diamond,color=black, very thick] table [x index = {0}, y index = {4}, col sep=comma] {data/stats.csv};
\addlegendentry{Total};
\end{axis}
\end{tikzpicture}
\vspace{-3mm}
\caption{Weekly Usage of \emph{dcc \texttt{-{}-}help} Over a Teaching Period}
\vspace{-3mm}
\label{figure:2}
\end{figure}

\autoref{figure:2} depicts weekly usage of the tool by our CS1 and CS2 students. In week one, 1,032 uses occurred, increasing to over 9,700 in the final week, demonstrating increasing popularity and adoption of the tool. On average, 1,077 unique students have used \emph{dcc \texttt{-{}-}help} every week, with a mean of 60 uses per student and a median of 38 uses per student. Overall, 93\% of our 2,565 CS1 and CS2 student cohorts have used the tool at least once during the teaching period. Usage peaked during weeks prior to major assessment due dates. So far, \emph{dcc \texttt{-{}-}help} has generated a total of 64,119 explanations in just ten weeks, with 49,866 compile-time, and 14,253 run-time explanations. We also recorded the time of day when students engaged the tool, and present brief reflections in \autoref{reflections}.

\begin{table}
  \caption{Review of Generative Explanations: Frequencies of "Yes" per Category and the Inter-Rater Reliability.}
  \label{table:1}
\begin{tabular}{lccc}
    \toprule
 Measure (\emph{n=400}) & \shortstack{ CT\\ (\emph{n=200})} & \shortstack{RT\\ (\emph{n=200})}  & \shortstack {Light's\\ $\kappa$}\\
\midrule
\textbf{Conceptually} accurate & 90\% & 75\% & 0.56\\
\textbf{No Inaccuracy} in solution & 78\% & 53\% & 0.45 \\ 
\textbf{Correctness} of response & 93\% & 66\% & 0.66 \\
\textbf{Relevance} of response & 92\% & 75\% & 0.45 \\
\textbf{Completeness} of response & 72\% & 39\% & 0.43 \\
\textbf{Solution} is provided & 48\% & 49\% & 0.73 \\
Response of \textbf{peer} quality & 28\% & 53\% & 0.45 \\
Response of \textbf{tutor} quality & 72\% & 45\% & 0.45\\
\bottomrule
\end{tabular}
\end{table}

\section{Discussion}\label{discussion}
Results are promising, and \textit{moderate} to \textit{substantial} index of agreement between the four reviewers validate the findings. We show that the use of LLMs to generate explanations of compiler error messages to augment compiler output in \emph{dcc \texttt{-{}-}help} is feasible when sufficient information such as error and stack trace is provided.

Our results indicate that LLM-generated explanations perform better at generating compile-time error explanations than run-time. We believe this is due to the increased context requirements at run-time---here, error explanations need to incorporate the \textit{state} of the program. %
We find that the LLM often provides correct conceptual explanation for these problems, but can fall short on smaller technical details, for instance occasionally mis-identifying bug lines. %
Despite this, we overall find that the explanations are helpful in solving code-related issues, with error descriptions and analysis commensurate with a junior member of our teaching team.

ChatGPT also had a propensity to disregard our prompts instructing it to not solve the problem and output solution source code. It is not entirely clear why this occurs, however, this was not pedagogically harmful. Future work may address this limitation using simple post-processing of results to remove code blocks, or prompt engineering for better LLM instruction.

\subsection{Reflections and Observations}\label{reflections}
We have seen overwhelming adoption of the tool amongst our students, with consistent growth in usage since its introduction shown in \autoref{figure:2}. We observed significant engagement in weeks preceding a major assessment (week 4 and 6), indicating students were turning to the tool. Overall, 47\% of \emph{dcc \texttt{-{}-}help} use occurs between the hours of 18.00 and 08.00, when teaching assistance is not readily available. This highlights a key advantage of the tool -- a method for student assistance outside staff office hours.

Cursory evaluation of performance in the invigilated, closed-book final exam in which \emph{dcc \texttt{-{}-}help} was not available indicates no significant performance loss compared to previous terms.

\subsection{Too much help?}
\label{sec:too-much-help}
Following introduction of \emph{dcc \texttt{-{}-}help} into our computing courses, questions and discussions naturally arose. Firstly, should \emph{dcc \texttt{-{}-}help} be made available to more students? Secondly, and on the flip side, does the tool in its current form provide too much assistance? While these are open questions, we acknowledge that regardless of our choices, students can access their own explanations using ChatGPT themselves. We believe that providing access to this tool outweighs the potential risks, especially as we designed the tool with limits (such as rate-limiting explanations with warnings), and removed its availability from the final exam.

\emph{dcc \texttt{-{}-}help} also affords us the opportunity to identify and discuss with our students limitations of generative AI tools, especially when the tools provide incorrect explanations. Ingraining a scepticism-first approach provides meaningful learning opportunities for students, ensuring that even when seeking assistance from \emph{dcc \texttt{-{}-}help} they should always have a clear understanding of their goals and code when debugging.
In many cases, we observed that the benefits of \emph{dcc \texttt{-{}-}help} were the clear and friendly language of the generative explanations, particularly at compile-time -- reinterpretations of cryptic error messages could lead students to an understanding of their errors and successful resolutions thereof.

Overall, the use of LLMs to generate contextual explanations for compiler errors provides a way to scaffold existing student support. The automated nature of the tool benefits the scale of learning that we face. Whilst at some stage, those scaffolds need to be removed \cite{Sweller2019CognitiveLater} and students need to understand compiler error messages on their own, they need not be expected to do this at the start of their learning journey -- and especially not in languages such as C. All cognitive focus should be on learning the programming language, as opposed to interpreting cryptic error messages provided by the compiler at the introductory level.  

\section{Limitations and Future Work}
Whilst the four reviewers assessed 15 errors together to form a consensus of interpretation of the task, different interpretations of the categories may have impacted the reliability due to the subjective nature. Still, we observe an overall \textit{moderate} to \textit{substantial} agreement in all classifications.

Currently, we have not assessed students' interpretations of generative explanations, nor the tool's efficacy in assisting students to understand and solve errors. Human research ethics approval has been obtained to explore this in the future.

Integrating OpenAI's \emph{gpt-3.5-turbo-0301} into DCC introduces costs which may impact scalability, however the API is affordable, costing \$104 USD to generate over 64,000 explanations.

Finally, OpenAI's newly released GPT4-based models claim improved performance in many contexts. Future work could explore the performance of these models in this context, and compare any benefits to the increased associated costs. Finally, there is considerable scope for alternative prompt formations, including, for example, the addition of compiler-tooling specifically to provide extra information for run-time error prompts to potentially improve the quality of run-time explanations.

\section{Conclusion}
Our open-source \emph{dcc \texttt{-{}-}help} tool presents a promising avenue for deploying Large Language Models (LLMs) to generate controlled, novice-focused compiler error messages directly in the development environment. We found LLM-generated explanations conceptually accurate in the majority of cases, and the popularity of the tool with our CS1 and CS2 cohorts at a large Australian university demonstrates student acceptance. Future work is required to evaluate the tool's usefulness from the student perspective. Initial reflections and observations, including a spike in the tool's use immediately before major assessment deadlines suggest students are choosing to continue engaging with the tool. Integrating generative explanations into the compiler allows us to scaffold novice students with contextual guidance the moment an error occurs, ensuring students can act upon the explanations. In our experiences, \emph{dcc \texttt{-{}-}help} has transformed the role of the compiler from a tool that continually repeats frustrating, unhelpful or cryptic error messages to a student-focused \textit{guide by the side} -- always available for those times when a student just needs a little bit of help.

\bibliographystyle{ACM-Reference-Format}
\balance
\bibliography{references_sasha, references_hammond_zotero}

\end{document}